\documentclass{article}

\usepackage{fullpage,amsmath,amssymb,latexsym}

\usepackage{graphicx}
\begin{document}
\title{Planetesimal Capture in the Disk Instability Model}
\author{Ravit Helled, Morris Podolak, and Attay Kovetz\\
Dept. of Geophysics and Planetary Sciences\\
Raymond and Beverly Sackler Faculty of Exact Sciences\\
Tel Aviv University\\
Tel Aviv, Israel 69978\\
helledra@post.tau.ac.il; morris@post.tau.ac.il; attay@post.tau.ac.il}
\maketitle \vskip 2cm
\noindent
%Pages: 20\\
%Tables: 3\\
%Figures: 6\\

%\newpage

%Proposed Running Head: Planetesimal Capture in the Disk
%Instability Model \vskip 4cm

%\noindent
%Editorial correspondence: Morris Podolak\\Dept. of Geophysics and Planetary Sciences\\
%Tel Aviv University\\
%Tel Aviv, Israel 69978\\
%Phone:+972-3-6408620\\ Fax:+972-3-6409282\\
%E-mail:morris@post.tau.ac.il
%\newpage

\begin{abstract}
We follow the contraction and evolution of a typical Jupiter-mass
clump created by the disk instability mechanism, and compute the
rate of planetesimal capture during this evolution. We show that
such a clump has a slow contraction phase lasting $\sim 3\times10^5$
years. By following the trajectories of planetesimals as they pass through the envelope of the protoplanet, we compute the cross-section for planetesimal capture at all stages of the protoplanet's evolution.  We show that the protoplanet can capture a large fraction of the solid material in
its feeding zone, which will lead to an
enrichment of the protoplanet in heavy elements.  The exact amount
of this enrichment depends upon, but is not very sensitive to the size and random speed of the planetesimals.
\end{abstract}

\bigskip

\newpage

\section{Introduction}

There are nearly 200 known extrasolar planetary systems.  Many of
these systems contain planets of several Jupiter masses.  For those
few systems that transit the stellar disk, the radius can be
measured, and most of these transiting planets appear to be gas
giants (see, e.g. Henry et al. 2000; Udalski et al. 2002; Pont et
al. 2004; Alonso et al. 2004). Currently, there are two main
theories for forming such planets. The core accretion theory
(Pollack et al. 1996) argues that a heavy element core is built up
by the accretion of planetesimals. As the core grows, its ability to
accrete gas from the surrounding disk increases. When the core is
sufficiently massive, there is a rapid accretion of such gas and a
giant planet is formed.  The big advantage of this model is that the
same basic mechanism of planetesimal accretion will form terrestrial
planets, Uranus and Neptune-sized intermediate planets, and giant
planets.  The disadvantages are that the core mass required to form
Jupiter is at the upper end of the mass estimated from interior
models (Saumon and Guillot 2004), and that the time to reach a
Jupiter mass is uncomfortably close to the upper limit estimated for
lifetime of the gas disk (Haisch, Lada, \& Lada 2001).  Both these objections can be ameliorated, if not entirely removed, if the accreted material does not sink entirely to the core (Pollack et al. 1996) and if the opacities are in fact lower than those estimated from interstellar grains (Podolak 2003; Hubickyj et al. 2005).

A competing mechanism for giant planet formation is a local disk
instability (Boss 1998a). This model suggests that under the right
conditions an instability can form in the protoplanetary disk. This
instability can lead to the creation of a self-gravitating clump of
gas and dust. Such clumps can contract to form giant gaseous
protoplanets (Boss 1997; Boss 1998b). This model has several
advantages, among them, that planets can be formed quickly, before
the nebular gas dissipates. Calculations by Boss (2000) show
that the unstable disk can break up into giant gaseous protoplanets
in $\sim10^3$ yrs. The instability occurs over dynamical timescales
(some tens of orbital periods), and the formation of clumps (future giant
gaseous protoplanets) can easily occur within the estimated lifetime
of most circumstellar disks (Haisch, Lada, \& Lada 2001).

One of the problems with the disk instability model is that the
planets formed by this mechanism start with a solar abundance of elements.
Observations as well as theoretical models, however, indicate that
Jupiter's envelope (and Saturn's as well) is enriched with heavy
elements (Young 2003; Saumon et al. 1995). The theoretical estimate
for the mass of heavy elements in Jupiter is $\sim 20 - 30
M_{\oplus}$ (Saumon and Guillot 2004).  A solar composition planet
of Jupiter's mass would be expected to have only $\sim 6 M_{\oplus}$
of heavy elements. This means that planets created by the disk
instability mechanism must have accreted this additional material
later, presumably as solids. In this paper we follow the evolution
of an isolated clump and compute the planetesimal accretion rate as
a function of time. The calculation is divided into three parts: the
computation of the evolution of an isolated clump, the computation
of the cross section for the two body interaction between the
protoplanet and a planetesimal, and the actual calculation of the mass
that is accreted. The details are presented in the following
sections.

\section{Planetary Evolution Code}
In our calculation we assume an isolated non-rotating clump. The
initial physical parameters of the body were chosen to fit the
expected initial conditions after the onset of the gravitational
instability (Boss 2002).  We assume that there are no external
influences on the body, such as solar heating or disk shear.  We
follow the evolution of this clump using a stellar evolution code
developed by one of us (Kovetz).

This code, which was originally
developed for stellar evolution studies, solves the standard
equations of stellar evolution:
\begin{equation}
\frac{\partial}{\partial m}\frac{4\pi}{3}r^3=\frac{1}{\rho},
\end{equation}\vspace{-0.8cm}
\begin{equation}
\frac{\partial p}{\partial m}=-\frac{Gm}{4\pi r^4},
\end{equation}\vspace{-0.8cm}
\begin{equation}
\frac{\partial u}{\partial t}+p\frac{\partial}{\partial t}\frac{1}{\rho}
=q-\frac{\partial L}{\partial m},
\end{equation}\vspace{-0.8cm}
\begin{equation}
\frac{\partial Y_j}{\partial t}=R_j-\frac{\partial F_j}{\partial m},\qquad
F_j=-\sigma_j\frac{\partial Y_j}{\partial m},
\end{equation}\vspace{-0.8cm}
\begin{equation}
\frac{\partial T}{\partial m}=\nabla\frac{\partial p}{\partial m},
\end{equation}
where $Y_j$ is the number fraction of the $j$'th species, related to
the mass fraction $X_j$ by $Y_j=X_j/A_j$, $F_j$ is the particle
flux of the $j$'th species, determined by the corresponding
coefficient of diffusion $\sigma_j$, $\nabla(r,m,L,\rho,T,Y)$ is the temperature
`gradient' $d\log T/d\log p$, determined by the Mixing Length Recipe (MLR),
and the remaining symbols are in standard notation.
At the center $r$, the energy flux $L$ and the $F_j$'s all
vanish. The star's surface is taken to be the photosphere.
Thus, the surface boundary conditions are
$L=4\pi r^2 \sigma_{SB} T^4$, $F_j=0$ and $\kappa p = Gm\tau_s/r^2$, where
$\tau_s$ is the optical depth of the photosphere, which we take to be unity.

The foregoing equations are replaced by implicit difference
equations, which are then solved numerically.
Instead of using a fixed grid of mass points, the code determines the mass
distribution by requiring the function
\begin{equation}
f=\Bigl({m\over M}\Bigr)^{2/3}-{c\over\log(p_c/p_s)}\log p
\end{equation}
to change by a constant increment between any two consecutive mass points.
With $c$ a numerical constant of order unity, the second-order difference
equation $d^2f=0$, which has the boundary conditions $m=0$ at the center
and $m=M$ at the surface %($M(t+dt)=M(t)+\dot M_{acc}dt$ if there is mass accretion)
, ensures equal steps of $\log p$,
except near the center, where it imposes equal steps of $m^{2/3}$.

In the absence of any nuclear (or chemical) change, the rates $R_j$
will all vanish, but the heating source $q$ may still be positive, e.g. when
(and where) accreted planetesimals are slowed by friction.

As noted above, convection is treated in accordance with MLR.
In order to avoid the difficulties connected with sudden, instantaneous
mixing in convective zones, the code regards convective mixing
as a diffusive process in a gas of density $\rho$, mean velocity
$v_c$ (supplied by MLR) and mean free path $l_c=1.5 H_p$. The artificial
convective diffusion coefficient $\sigma_c=(4\pi r^2\rho)^2 v_c l_c$
(the same for all species) of course vanishes
outside the convective zones. Any real diffusion, or settling, process can then be
easily incorporated by adding actual diffusion coefficients.

The equation of state tables were kindly provided by D. Saumon
and are based on Saumon et al. (1995), supplemented at low pressures by our own equation of state,
appropriate for a weakly interacting (Debye approximation) mixture
of gases.  The opacity tables were kindly provided by P. Bodenheimer, based on the work of Pollack
et al. (1989).  They include both gas and grain opacity, the latter being based on the size distribution relevant for interstellar grains.

We were able to find an initial quasi-static model
of a Jupiter mass object, similar to the initial clump in the model
of Boss (2002), based on a preliminary model, also provided by
Bodenheimer.  Table 1 gives the initial parameters of the starting model.
The pressure and temperature profiles for the initial model are shown in fig. 1.

\begin{table}[ht]
\centering
\begin{tabular}{|l|c|c|}
\hline
Properties of the initial Model&\\
\hline
Radius (cm)&$7.14\times 10^{12}$\\
Effective temperature (K)&26.3\\
Central Temperature (K)&$3.51\times 10^{2}$\\
Photospheric density (g cm$^{-3}$)&$1.64\times 10^{-11}$\\
Central density (g cm$^{-3}$)&$2.53\times 10^{-8}$\\
Photospheric pressure(dyne cm$^{-2}$)&$1.55\times 10^{-2}$\\
Central pressure(dyne cm$^{-2}$) &$3.17\times 10^{2}$\\
\hline
\end{tabular}
\caption{Properties of the initial model} \label{tab:1}
\end{table}

\begin{figure}[ht]
    \centering
    \includegraphics[width=3.5in]{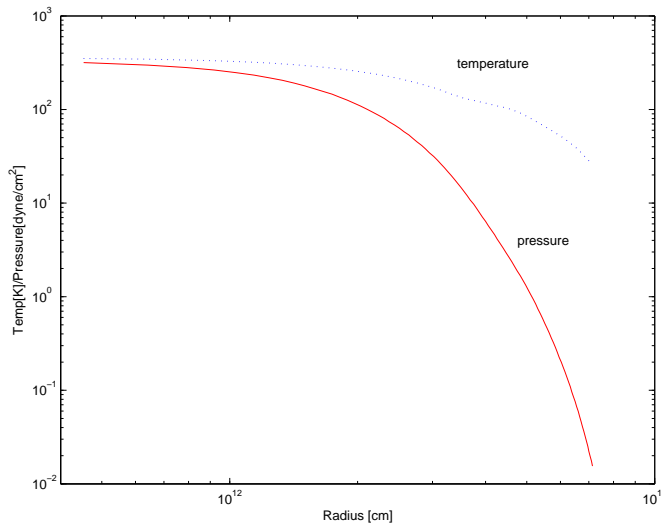}
    \caption[8pt]{ Pressure and temperature as a function of radius for the initial model}
\end{figure}

Fig. 2 shows evolution of the protoplanet's radius, central temperature, and central pressure with time for the first
$3\times 10^5$yr.  At this point the central temperature reaches
$\sim 2000$ K and the molecular hydrogen dissociates sufficiently so
that the resultant energy sink triggers a rapid collapse of the
body. The evolutionary track is similar to one calculated by
Bodenheimer at al. (1980), for a $1M_J$ protoplanet with an initial
central density of $10^{-9}$g cm$^{-3}$ and an initial central
temperature of $100$ K. For comparison our $1M_J$ protoplanet initially
has an central density of $\sim 2\times10^{-8}$g cm$^{-3}$, a radius of $0.5$ AU,
and a central temperature of $\sim350$ K.  The evolution of
the body was computed assuming no planetesimal capture, so that the composition of the protoplanet was constant.

\begin{figure}[ht]
    \centering
    \includegraphics[width=3.5in]{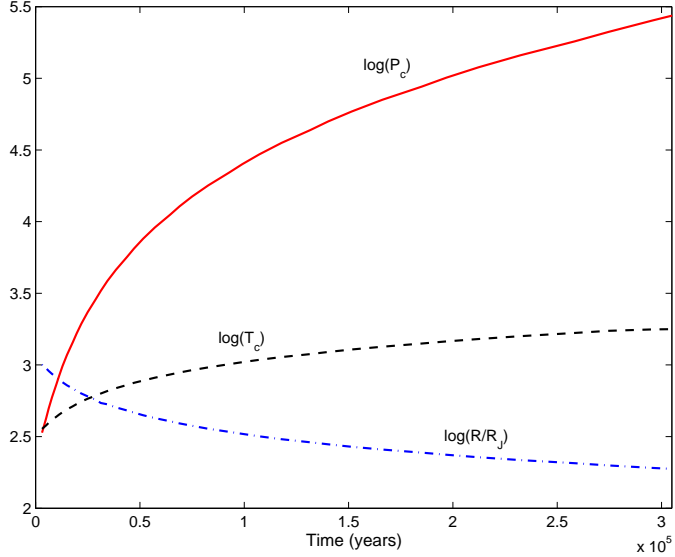}
    \caption[8pt]{Evolution with time of the central pressure (solid curve), the central temperature (dashed curve), and the radius of the protoplanet in Jupiter radii (dash-dot curve). }
\end{figure}

\section{Two-Body Capture Cross Section}

Once the structure of the planet is determined, one can, in
principle, compute its capture cross-section.  If this is known, the
time-dependent rate of mass increase due to the capture of
planetesimals is given by (Safronov 1969)
\begin{equation}
\frac{dm}{dt}=\Sigma (t)\sigma \Omega
\end{equation}
where $\Sigma (t)$ is the protoplanet's cross section for
planetesimal capture, $\sigma $ is the surface density of solid
material, and $\Omega $ is the orbital frequency of the
protoplanet.

To compute the capture cross section at each stage of the evolution,
we followed the trajectory of a planetesimal as it passed through
the protoplanetary atmosphere.  At each step of the trajectory we computed the
motion of the planetesimal in response to the gas drag and the
gravitational forces, assuming that the gravitational interaction is
only two-body.  The effects of heating and ablation of the
planetesimal as it passed through the protoplanetary envelope were
included.  We also allowed for the possibility of planetesimal
fragmentation. This will occur when the pressure gradient across the
planetesimal exceeds the material strength of the body, and
the planetesimal is small enough so that self-gravity cannot counteract the disruptive effect of the pressure
gradient (Pollack et al. 1979).  Further details of the code for
this calculation are described in Podolak et al. (1988).

We computed trajectories for planetesimals with successively larger
impact parameters with the protoplanet in order to determine the
largest impact parameter for which planetesimal capture was
possible. For this trajectory the planetesimal's closest distance of
approach to the center of the protoplanet was defined as the
effective radius for capture $R_{eff}$.
Of course $R_{eff}$ depends on the (instantaneous) protoplanetary
density distribution, as well as on the size, composition and random
velocity of the planetesimals.

The planetesimals were assumed to be composed of a mixture of ice,
silicates and CHON with a density of 2.8 g cm$^{-3}$. In what
follows, this mixture of ice, rock and CHON will be referred as "ice
+ rock". For comparison we also ran some cases with pure rock
planetesimals with a density of 3.4 g cm$^{-3}$.  The
planetesimal material properties are given in Table 2.

\begin{table}[ht]
\centering
\begin{tabular}{||l|c|c||}
\hline
Physical parameter&Ice&Rock\\
\hline
$\mu$ (g/mole)&18&50\\
$\rho$ (g cm$^{-3}$)&1&3.4\\
$Log_{10}P_{vapor}$ (dynes cm$^{-2}$)&$\frac{-2104.2}{T}+11.5901$&$\frac{-24605}{T}+13.176$\\
 \hline
\end{tabular}
\caption{Properties of the planetesimal material} \label{tab:2}
\end{table}

The effective radius for planetesimal capture will depend not only on the size and composition of the planetesimals, but also on their random velocity. Following Pollack et al. (1996) we have
taken the random velocity of the planetesimal far from the
protoplanet to be 1 km s$^{-1}$ in our baseline model.  We have also considered planetesimals with random velocities of 2 km s$^{-1}$. This is already 15\% of the keplerian velocity at Jupiter's heliocentric distance, and it is unlikely that the random velocities can be stirred up much higher than this in the short time before most of the planetesimals are captured.
Fig. 3 shows the ratio of the effective radius for planetesimal capture to
the actual radius of the protoplanet for
planetesimals with radii of 1, 10, and 100 km composed of ice +
rock and for the two values of the random velocity we considered. As can be seen from the
figure, for the higher random velocity the effective radii decrease since
it is harder to slow down and capture the planetesimals (especially
those with larger radii). Similarly, planetesimals with lower
velocities than in our baseline case will be captured more
efficiently.

\begin{figure}[ht]
    \centering
    \includegraphics[width=4.5in]{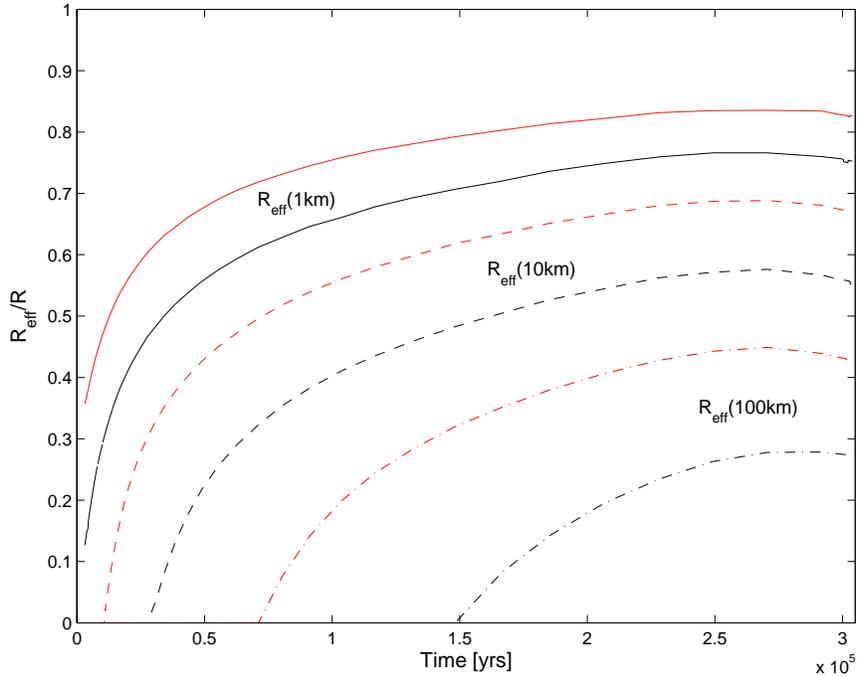}
    \caption[8pt]{Ratio of the effective radius of the planet to the actual radius as a function of time for ice + rock planetesimals with radii of 1 km (solid curves), 10 km (dashed curves), and 100 km (dash-dot curves).  For each case two curves are shown.  The upper curve is for a random velocity of 1 km s$^{-1}$ and the lower curve is for a random velocity of 2 km s$^{-1}$.}
\end{figure}
The additional heating supplied by the influx of planetesimals is
not expected to affect the course of the planet's evolution
significantly.  The radius of the protoplanet should increase as a
result, and its energy will increase as well.  The temperature (both
central and effective) will decrease in accordance with the well
known negative heat capacity of a self gravitating object. The
decrease of the effective temperature $T_e$ is more than offset by the increase of $R$,
so that the luminosity $L=4\pi R^2\sigma_{SB}T_e^4$ increases. As a result, the Kelvin-Helmholz
time, which is inversely proportional to the product $RL$ will actually
be somewhat shorter.  Detailed simulations confirm this simple
picture:  Initially the radius for the case with planetesimal heating was 30\% larger than for the case without planetesimal heating. After some $10^4$ yr the radius was only 15\% larger, and by $10^5$ yr the
difference was 3\%.  The effect of the planetesimals accreted during the evolution on the protoplanetary mass was negligible.

\section{Mass Accretion}
As can be seen from fig. 3, $R_{eff}$ is initially much smaller than
the actual physical radius of the protoplanet.  This is because the
outer layers of the protoplanet have such low density that the
planetesimal cannot be sufficiently slowed until it enters deeper
into the protoplanetary atmosphere. As a result, only those
planetesimals hitting the protoplanet with relatively small impact
parameters will be captured.

As the protoplanet contracts, however, the density throughout most
of its volume increases and the drag force becomes more important.
Initially the effective radius increases, but once the interior
densities are high enough, and the protoplanet continues to
contract, the effective radius decreases again.  The point at which
the gas density becomes high enough to efficiently capture
planetesimals depends on their size, and as can be seen from the
figure, the turn around point is later for larger planetesimals.  In
addition, the larger planetesimals cannot be captured at all
initially.  Planetesimals with radii of 10 and 100 km cannot be
captured during the first $\sim1\times10^4$ and $\sim7.5\times10^4$
years, respectively.  Although this means that the time for
planetesimal capture is reduced significantly for these cases, we
will show that the time remaining is still sufficient to accrete most of
the solids in the feeding zone.

There is a complication due to the fact that encounters between
planetesimals and the protoplanet actually take place in the
presence of the Sun, and should really be treated as three-body
encounters.  In this case, the two-body cross section needs to be
multiplied by a gravitational enhancement factor: $\Sigma = F_g\pi
R_{eff}^2$ which corrects for the three-body effects.  Greenzweig
and Lissauer (1990; 1992) found expressions for $F_g$ based on fits
to numerical simulations of the planetesimal orbits.  For the
conditions studied here, their fits give an $F_g$ which is typically
several hundred, and can be as large as $10^3$.  However, the fits
of Greenzweig and Lissauer were done for protoplanets small compared
to the Hill sphere radius.  In our case, the clump has an initial
radius which is essentially equal to the Hill sphere radius.  As a
result, it is not clear that the numerical results of Greenzweig and
Lissauer apply in this case.

An alternative is to use the two-body gravitational cross section,
$\pi b^2$ where $b$ is the largest impact parameter for which a
planetesimal is captured, and is computed using the planetesimal trajectory code described above.  This however, neglects any influence due to
the Sun.  Fig. 4 shows the difference between $R_{eff}$ and $b$ for
different cases.  As can be seen from the figure, the difference is,
for the most part, a factor of a few.  We will here use the most
conservative estimate of the cross section, and take it equal to
$\Sigma = \pi R_{eff}^2$.  This will overestimate the time required
to accrete the planetesimals in the protoplanet's vicinity.

\begin{figure}[ht]
    \centering
    \includegraphics[width=4.5in]{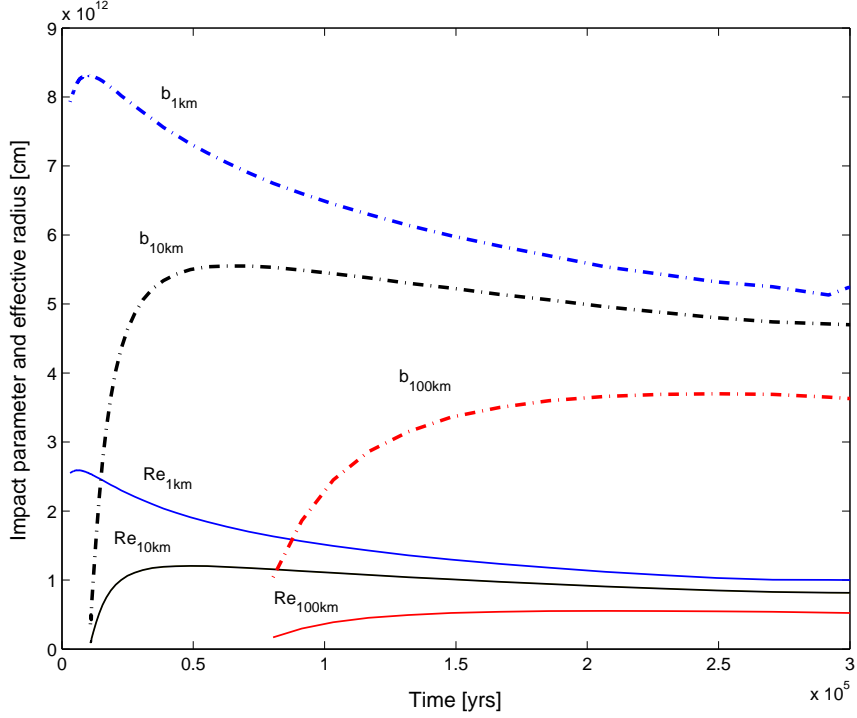}
    \caption[8pt]{Comparison of the effective radius (curves labeled "Re") and the impact parameter (curves labeled "b") for different planetesimal sizes as a function of time.}
\end{figure}

\newpage
The total mass of solids in the planet's feeding zone is given by
\begin{equation}
M_{disk}=\pi (a_{out}^2-a_{in}^2)\sigma (t)
\end{equation}
where $a_{out}$ and $a_{in}$ are the outer and inner radii of the
feeding zone, respectively. Again, following Pollack et al. (1996)
we assume $\sigma =10$ g cm$^{-2}$ in the region of Jupiter's
formation, and that the planetesimals are spread uniformly on either
side of the orbit to a distance $a_f$, which depends on the
eccentricity and inclination of the planetesimal orbits, as well as
the Hill sphere radius of the protoplanet (see Pollack et al. for
details).  For conditions similar to those assumed by Pollack et al.
$a_f = 2.2\times 10^{13}$ cm, $a_{out}=a+a_f$ and $a_{in}=a-a_f$.
The mass of solids in Jupiter's feeding zone using these values is
$2.2\times10^{29}$ g $\sim 38 M_{\oplus}$.  This is roughly twice
the estimated amount of heavy elements in Jupiter (Saumon and
Guillot 2004), so that even if some of the planetesimals are
scattered out of Jupiter's feeding zone, the protoplanet should still
manage to capture enough material to match theoretical models of
Jupiter.

We can now use eq. 1 to compute the rate of mass capture by the
protoplanet. We consider three sizes of planetesimal: 1, 10, and 100
km.  Since a clump is expected to start forming relatively quickly,
it is likely that the planetesimals were in the small end of this
range.  Fig. 5 shows the mass of accreted material as a function of
time for the three planetesimal sizes we considered.  In all cases
the planetsimals are assumed to be composed of a mixture of rock and
ice in solar proportions.  For the purpose of computing the drag
force and the resultant ablation of the planetesimal, we view it as
composed of rocky material embedded in a matrix of ice.  When the
temperature is high enough for the ice to ablate, a proportionate
amount of rock is lost together with it.  For such planetesimals,
the heating and mass loss are similar to those for pure ice, and
they are easily captured.

\begin{figure}[ht]
    \centering
    \includegraphics[width=4.5in]{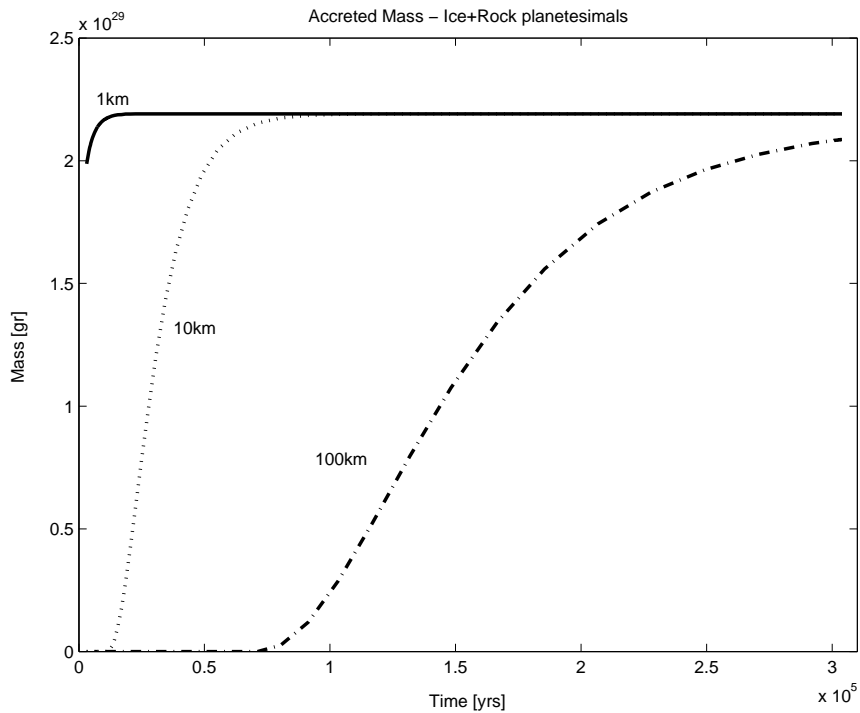}
    \caption[8pt]{Captured mass for the three sizes of planetesimals made of ice and rock vs. time.}
\end{figure}

1 km planetesimals can be captured almost immediately, while 10 km
planetesimals can only be captured after the protoplanet has
contracted for some tens of thousands of years.  100 km
planetesimals require $\sim 10^5$ yr of  evolution before the
average density of the protoplanet becomes high enough to capture
them.  In the first two cases all of the available planetesimal mass
is captured.  In the case of 100 km planetesimals some 90\% of the
available mass is captured.

Fig. 6 shows the same calculation for the assumption of pure rock
planetesimals.  As expected, the protoplanet must contract more than
for the equivalent case for planetsimals composed of a mixture of
ice and rock, but the difference is not large.  For both 1 and
10 km planetsimals all of the available mass is captured within
$\sim 10^5$ yr.  For the 100 km planetsimals $\sim 80\%$ of the mass
is captured before the protoplanet enters the rapid contraction
phase.  Since it is likely that the planetesimals will have some
distribution is size, some of the mass will be in smaller bodies,
and the fraction of solid material that will be captured will be
even closer to 100\%.  Thus, assuming that none of the material is
gravitationally scattered out of the feeding zone, a contracting
protoplanet is expected to accrete most of the mass in its feeding zone
well before it evolves to the rapid contraction stage.

\begin{figure}[ht]
    \centering
    \includegraphics[width=4.5in]{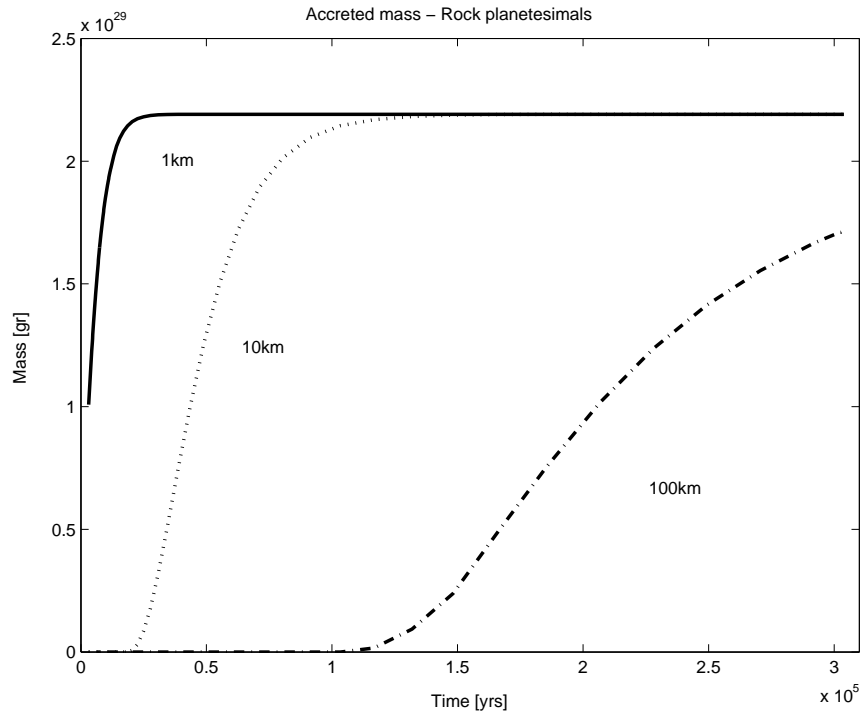}
    \caption[8pt]{Captured mass for the three sizes of planetesimals made of pure rock vs. time.}
\end{figure}

The total mass of solids in the feeding zone is a direct consequence
of the assumed solid surface density. Using higher values of
$\sigma$ increases the total mass of solids in the accretion zone.
As it is usually presented, the disk instability model requires a greater surface
density (by $\sim 150\%$) for an instability to form. We therefore repeated our calculation
using $\sigma$ as $15$ g cm$^{-2}$. In this case, the total mass of
the planetesimals in the feeding zone is $\sim 3.3\times10^{29}$ g $\sim 55M_{earth}$
and, as in the precvious case, essentially all the
planetesimals are captured. We summarize our results in table 3.
Here the "capture time" is the time required to capture essentially
all of the planetesimals in the feeding zone.  If this time is
longer than $\tau_{col}=3.04\times 10^5$ yr, the time for the
protoplanet to reach the rapid collapse stage, then $\tau_{col}$ is
given in the table as the capture time.  As can be seen, all 1 km
and 10 km planetesimals can be captured within a relatively
short time. Things are slightly different with the 100 km
planetesimals: here the accreted mass is $\sim 95\%$ of the solids
for a random velocity, $V_{\infty}$ of $10^5$ cm s$^{-1}$.  For
$V_{\infty}=2\times10^5$ cm s$^{-1}$, less than half of the material
is captured.

\begin{table}[ht]
\centering
\begin{tabular}{||c||c|c||c|c||c|c||}
\hline\hline &\multicolumn{2}{c||}{$\sigma =10$ g/cm$^2$,
$V_{\infty}=1$ km/s} &\multicolumn{2}{c||}{$\sigma =10$ g/cm$^2$,
$V_{\infty}=2$ km/s} &\multicolumn{2}{c||}{$\sigma =15$ g/cm$^2$,
$V_{\infty}=1$ km/s}
\\
\hline
 $r_p$ (km) &Time (years)& $M_{captured}$(g) & Time (years) & $M_{captured}$(g) & Time (years) & $M_{captured}$(g)\\
\hline\hline
$1$& $2.09\times10^3$ &  $2.19\times10^{29}$& $4.89\times10^4$ & $2.19\times10^{29}$& $3.66\times10^3$& $3.29\times10^{29}$\\
\hline
$10$& $1.52\times10^4$&  $2.19\times10^{29}$ & $2.99\times10^5$&  $2.19\times10^{29}$ &$1.63\times10^4$&$3.29\times10^{29}$\\
\hline
$100$ &$3.04\times10^5$&  $2.09\times10^{29}$ &  $3.04\times10^5$ & $9.88\times10^{28}$& $3.04\times10^5$ & $3.13\times10^{29}$ \\
\hline
\end{tabular}
\caption{The capture time and accreted mass for different cases.
All cases correspond to ice+rock planetesimals} \label{tab:3}
\end{table}

\section{Summary and Conclusions}

We present an evolutionary model of an isolated giant gaseous protoplanet such as might be formed from a clump caused by an instability in the solar nebula. Assuming that such a clump survives and becomes a protoplanet, we calculate the rate at which it would capture planetesimals. We considered three planetesimal sizes and two types of materials, ice + rock, and pure rock. The total mass of solids that is captured depends on the size
distribution of bodies in the accretion zone and on their composition. 100 km-sized planetesimals can not be captured during the first
$\sim10^5$ years of the evolution. This is due to the fact that
until that time the density and temperature throughout most of the volume of
the protoplanet are too low. Bodies with sizes of 1 km or less can be
accreted starting at times shortly after the clump's creation.  Icy bodies are captured more easily than pure rock planetesimals.

In the case of higher random velocities, the
capture efficiency decreases and the time for capturing the same
amount of solids is longer, but even with random velocities as high as 15\% of the
 keplerian velocity, more than 16 $M_{\oplus}$ of heavy elements can be captured. For a higher surface density (bigger amount of solids in Jupiter's feeding zone), the total mass
of heavy elements that is captured increases proportionately.
As can be seen
from table 3 the accreted mass varies by a factor of two when the
random velocity of the 100 km planetsimals is increased from 1 to
2 km s$^{-1}$.  Since the latter is already 20\% of Jupiter's
orbital speed, it is unlikely that the random velocities will be
significantly higher than this, especially since most of the planetesimals
are captured before they can be highly stirred.  Even in this case
some 50\% of the mass in the feeding zone is captured.  If, as is
likely, there is a distribution of planetesimal sizes, with 100 km
being at the high
end, then the fraction of feeding zone mass that is captured will be even higher.

It has been suggested (e.g. Weissman 1986) that planetesimals are actually loosely bound "rubble piles".  If this is indeed the case, they would be held together even more weakly than our ice + rock planetesimals.  Such planetesimals would be captured more easily that those composed of ice + rock, and again, the values we have presented for the mass captured would be a lower limit.

Of course all this assumes that the planetesimals remain in the feeding zone of the protoplanet
 and are not scattered out (see, e.g., Hahn and Malhotra 1999). It should be noted, however, that simulations of planetesimal scattering treat the planets as point masses, and do not include the gas drag on the planetesimal as it goes through the protoplanet's extended envelope.  The energy loss which results from the gas drag will act to inhibit scattering.  In addition, the larger cross section for capture that an extended protoplanet presents, will also significantly increase the ratio of
 captured to ejected planetesimals.  Thus, in the simulations of Hahn and Malhotra, for example,
 significant mass ejection requires times much longer than the $3\times 10^5$ years required for
the protojupiter to accrete most of the mass in its feeding zone.  Therefore, if giant planets
are created by the disk instability mechanism, there is a good chance that many of the planetsimals
will be accreted before they are scattered out of the
feeding zone.  As a consequence, these protoplanets can collect a substantial amount of
solids during their evolution and end up with a significantly
non-solar composition. This finding removes an important objection to the disk instability
model as a mechanism for giant planet formation.

Our model assumes a non-rotating clump, in reality, the clump is
likely to have some angular momentum.
This might act to slow the contraction of the clump in its later stages, and lead to the formation of a subdisk.  Planetesimals in this subdisk could then provide the material for forming both regular and irregular satellites.
It is also interesting to speculate on the possibility of observational
verification of our results.  Some models of extrasolar planets
whose densities are known seem to indicate non-solar compositions
(e.g., Bodenheimer et al. 2001). The extent of the enhacement of
the non-solar component may provide limits on the proposed
accretion mechanism. However, both the fits to the observations
and the models we have computed, have, as yet, too many free
parameters,
to allow useful constraints.

There are a number of issues which may affect the conclusions
presented above. In the first place, the captured planetesimals will
contribute high-Z material to the gaseous envelope, and change its
composition.  This will also affect the opacity of the envelope. Since much of this additional
material will end up in the convective region of the planet, this change in opacity should not
have a large effect on its subsequent evolution.  In addition, there will be some small effect
due to the additional mass of the planetesimals themselves.  Finally,
the high-Z material which remains in the envelope in the form of
grains will sediment towards the center of the protoplanet leading
to the formation of a core, and this too should affect the final
evolution of the body.  We hope to address these issues in more detail in future
work.

\section{Acknowledgments}

Support for this work was provided by a grant from the Israel Science Foundation.

\newpage

\section{References}

Bodenheimer, P., Lin, D.N.C. and  Mardling, R.A., 2001, On the
Tidal Inflation of Short-Period Extrasolar Planets. ApJ, 548,
466B-472.\\\\Boss, A.P. 1997, Giant planet formation by
gravitational instability. Science, 276, 1836-1839.\\\\Boss, A.P.
1998, Evolution
of the Solar Nebula. IV. Giant Gaseous Protoplanet Formation. ApJ, 503,923-937.\\
\\Boss, A.P. 2000, Formation of extrasolar giant planets: core accretion or
disk instability? , Earth, Moon and Planets, 81, 19-26.\\\\Boss,
A.P. 2000, Possible rapid gas giant planet formation in the solar
nebula and other protoplanetary disks. ApJ, 536,
L101-L104.\\\\Boss, A.P. 2001, Gas giant protoplanet formation:
Disk instability models with thermodynamics and radiative
transfer. ApJ, 536, 367-373.\\\\Boss, A.P. 2002, Evolution of the
Solar Nebula. V. Disk instabilities with varied thermodynamics.
ApJ, 576, 462-472.\\\\Greenzweig, Y. and Lissauer, J.J. 1990,
Accretion Rates of Protoplanets. Icarus 87, 40-77.\\\\Greenzweig,
Y. and Lissauer, J.J. 1992, Accretion Rates of Protoplanets II.
Gaussian Distributions of Planetesimal Velocities. Icarus 100,
440-463.
\\\\Guillot, T.,
Chabrier, G. ,Morel,P., \& Gautier, D. 1994, Nonadiabatic models
of Jupiter and Saturn. Icarus, 112, 354-367.\\\\Guillot,T.,
Gautier,D., \& Hubbard,W.B. 1997, Note: New Constraints on the
Composition of Jupiter from Galileo Measurements and Interior
Models, Icarus, 130, 534-539.\\\\Hahn, J. M. \& Malhotra, R. 1999,
Orbital evolution of planets embedded in a planetesimal disk.
Astron. J., 117, 3041-3053.\\\\Haisch, K.R. Lada, E.A., and Lada,
C.J., 2001,Circumstellar Disks in the IC 348 Cluster, AJ, 121,
2065-2074.\\\\ Hubickyj, O., Bodenheimer, P. and Lissauer,J.J.,
2005, Accretion of the gaseous envelope of Jupiter around a 5 10
Earth-mass core, Icarus,179,415-431.  \\\\Kenyon, S. J. and Luu,
J. X., 1998, Accretion in the early Kuiper Belt I. Coagulation and
velocity evolution. Astron. J. 115, 2136-2160.\\\\Podolak, M.
2003, The contribution of small grains to the opacity of
protoplanetary atmospheres.  Icarus 165, 428-437.\\\\Podolak, M.,
Pollack, J.B., and Reynolds, R.T. 1988, Interactions of
planetesimals with protoplanetary atmospheres. Icarus 73,
163-179.\\\\Pollack, J. B., Burns, J. A., and Tauber, M. E. 1979,
Gas drag in primordial circumplanetary envelopes: A mechanism for
satellite capture. Icarus, 37, 587-611.\\\\Pollack, J.B.,
Hubickyj, O.,Bodenheimer, P., Lissauer, J.J. ,Podolak, M., \&
Greenzweig, Y. 1996, Formation of the giant planets by concurrent
accretion of solids and gas. Icarus, 124, 62-85.\\\\Safronov, V.
S. 1969, Evolution of the Protoplanetary Cloud and the Formation
of the Earth and Planets. Nauka, Moscow. English translation: NASA
TTF-667. 1972.
\\\\Saumon, D., Chabrier, G., and Van Horn, H.M. 1995, An equation
of state for low-mass stars and giant planets. ApJ, 99,
713-741.\\\\Saumon,D. and Guillot,T. 2004. Shock Compression of
Deuterium and the Interiors of Jupiter and Saturn. ApJ, 609,
1170-1180.
\\\\Stevenson, D.J 1982. Formation of the giant planets. Planetary
and Space Science, 30, 755-764.\\\\Storm, S.E., Edwards, S., \&
Skrutskie, M.F. 1993, in Protostars and Planets III, ed. Levy,
E.H. \& Lunine, J.J (Tucson: Univ. Arizona Press),
837.\\\\Weissman, P. 1986.  Are cometary nuclei primordial rubble
piles? Nature 320, 242-244. \\\\Young, R.E. 2003.  The Galileo
probe: how it has changed our understanding of Jupiter. New
Astronomy Reviews, 47, 1-51.
\end{document}